\documentclass [12pt]{article}
\usepackage {amssymb}
\usepackage {amsmath}
\usepackage[dvips]{epsfig}

\begin{document}

\begin{center}
\Large\textbf{MICROSCOPIC MECHANISM RESPONSIBLE FOR
RADIATION-ENHANCED DIFFUSION OF IMPURITY ATOMS}
\\[2ex]
\normalsize
\end{center}

\begin{center}
\textbf{O. I. Velichko}
\end{center}

\begin{center}
\bigskip


{\it E-mail address (Oleg Velichko):} velichkomail@gmail.com

\end{center}

\textit{Abstract.} Modeling of radiation-enhanced diffusion of
boron and phosphorus atoms during irradiation of silicon
substrates respectively with high- and low-energy protons was
carried out. The results obtained confirm the previously arrived
conclusion that impurity diffusion occurs by means of the
``impurity atom -- intrinsic point defect'' pairs and that the
condition of the local thermodynamic equilibrium between
substitutional impurity atoms, nonequilibrium point defects
created by irradiation, and the pairs is valid. It is shown that
using radiation-enhanced diffusion, one can form a special
impurity distribution in the semiconductor substrate including
retrograde profiles with increasing impurity concentration into
the bulk of a semiconductor. The calculations performed give clear
evidence in favor of further investigation of various doping
processes based on radiation-enhanced diffusion, especially the
processes of plasma doping, to develop a cheap method for
formation of specific impurity distributions in the near surface
region.
\bigskip

{\it PACS:} 88.40.jj, 81.40.Wx, 66.30.Dn, 61.72.U-, 61.72.sh

{\it Keywords:} silicon; boron; phosphorus; radiation-enhanced
diffusion; modeling

\section{Introduction}

The increase in the cost of traditional energy resources has been
a very important tendency during the past years. Therefore,
investigation and implementation of alternative renewable energy
sources are particularly urgent now. One of the promising
alternative energy sources are solar cells. It is worth noting
that the best crystalline silicon photovoltaic modules are 5 \%
more efficient than the best modules based on polysilicon films
\cite{Saga-10,Green-12}. Production of solar elements includes
doping of silicon with impurity of \textbf{\textit{p}}- and
\textbf{\textit{n}}-types of conductivity
\cite{Haarahiltunen-09,Talvitie-11,Sepeai-12,Normann-13}. During
doping, simultaneously with the high concentration diffusion of
dopant atoms, the gettering of undesirable impurities such as iron
occurs \cite{Haarahiltunen-09,Talvitie-11} that results in
improvement of the electrophysical parameters of semiconductor
devices. It is evident that the production of solar elements could
be successful when a high value of energy conversion was achieved
at a low cost of technological processes. Therefore, the
introduction of impurity atoms into the silicon lattice by means
of the low temperature doping from a gas discharge plasma (plasma
immersion ion implantation or PIII)
\cite{Pinter-99,Chu-00,Chu-04,Torregrosa-05} is rather promising.
it is worth noting that thermal \cite{Pinter-99,Chu-00} or laser
annealing \cite{Torregrosa-05} is applied for electrical
activation of the induced impurity atoms. During PIII, the silicon
substrate can be heated up. On the other hand, a strictly assigned
elevated temperature of it can be provided by a special heater. In
such a case, the distribution of impurity atoms will be determined
by both the low energy impurity implantation and low temperature
radiation-enhanced diffusion (RED) that occurs due to the
generation of a great amount of nonequilibrium point defects
during plasma treatment.

One of the first experimental investigations of plasma doping has
been carried out in the paper of Strack et al. \cite{Strack-63}.
As follows from the numerous experimental data concerning
radiation-enhanced diffusion that the redistribution of the
impurity atoms previously introduced also occurs during the
implantation of hydrogen ions
\cite{Baruch-75,Baruch-1975,Baruch-77,Akutagawa-79,Lucas-80,Loualiche-82,Kozlovski-00},
silicon ions \cite{Venezia-04}, or heavy ions of inert gases
\cite{Holldack-87,Kachurin-88} in silicon substrates having an
elevated temperature. The experimental results obtained have shown
that the radiation-enhanced diffusion takes place at very low
temperatures insufficient for the ordinary thermal diffusion. For
example, it was reported in \cite{Venezia-04} that the RED of
boron atoms is observed below 100 $^{{\rm o}}$C. It means that for
local ion implantation the radiation-enhanced diffusion occurs
only in the local region with an increased concentration of point
defects and there is no redistribution of impurity atoms in the
other regions of the semiconductor. This feature of
radiation-enhanced diffusion can be very useful for the production
of various semiconductor devices including solar elements. Let us
consider the characteristic features of impurity transport
processes under conditions of radiation-enhanced diffusion.

\section {The equation describing the radiation-enhanced
diffusion of impurity atoms}

The equation for impurity diffusion due to the formation,
migration, and dissociation of the ``impurity atom -- vacancy''
and ``impurity atom -- silicon self-interstitial'' pairs has been
obtained in \cite{Velichko-84}. This equation takes into account
different charge states of all mobile and immobile species and the
drift of the pairs in the built-in electric field, although only
the concentrations of neutral point defects are involved in the
explicit form. It is supposed that nonuniform distributions of
nonequilibrium point defects including defects in the neutral
charge state can be formed and the mass action law is valid for
the pairs, substitutionally dissolved impurity atoms, and
self-interstitials or vacancies.

In the case of low impurity concentration $C \le n_{i}$, the
equation obtained can be presented in the form

\begin{equation} \label{DifEqLow}
{\frac{{\partial \,C}}{{\partial \,t}}} = D_{i}^{E} \,\Delta
\,\,\left( {\tilde {C}^{V\times} C} \right) + D_{i}^{F} \,\Delta
\,\,\left( {\tilde {C}^{I\times} C} \right) \, .
\end{equation}

Here $C$ is the concentration of substitutional impurity atoms;
$n_{i}$ is the intrinsic carrier concentration; $D_{i}^{E} $ and
$D_{i}^{F} $ are the diffusivities of impurity atoms in intrinsic
silicon due to the ``impurity atom -- vacancy'' and ``impurity
atom -- silicon self-interstitial'' pairs, respectively; $\tilde
{C}^{V\times}$ and $\tilde {C}^{I\times}$ are the concentrations
of vacancies and silicon interstitial atoms in the neutral charge
state normalized to the thermally equilibrium concentrations of
these species $C^{V\times}_{eq}$ and $C^{I\times}_{eq}$,
respectively.

In a number of doping processes only one kind of defect is
involved in diffusion of the main fraction of impurity atoms. For
example, during a low-temperature oxidation of the surface,
silicon self-interstitials are the dominating defects in the
silicon crystal \cite{Antoniadis-78} and diffusion of boron,
arsenic, and phosphorus occurs due to the interaction with these
interstitial atoms. On the other hand, it follows from
experimental data that antimony atoms diffuse due to the
interaction with vacancies \cite{Faney-84}. Then, Eq.
(\ref{DifEqLow}) can be presented in a simplified form as

\begin{equation} \label{DifEq1}
{\frac{{\partial \,C}}{{\partial \,t}}} = D_{i} \,\Delta
\,\,\left( {\tilde {C}^{\times} C} \right) \, ,
\end{equation}

\noindent where $\tilde {C}^{\times}$ is the concentration of the
neutral defects responsible for the impurity diffusion normalized
to the thermally equilibrium concentrations of this species
$C^{\times}_{eq}$.

The equation obtained retains the basic character of the origin
equations from \cite{Velichko-84}, namely, the ability to describe
segregation of impurity atoms, including the ``uphill'' impurity
diffusion. Indeed, we shall present Eq. (\ref{DifEq1}) in the
following form:

\begin{equation} \label{DifEq1Drift}
{\frac{{\partial \,C}}{{\partial \,t}}} = D_{i} \,\nabla \,\left(
{\tilde {C}^{\times} \nabla \,C} \right) + D_{i} \,\nabla
\,{\left[ {\left( {\,\nabla \,\tilde {C}^{\times} } \right)\,\,C}
\right]} \, .
\end{equation}

It is clearly seen from Eq. (\ref{DifEq1Drift}) that depending on
the gradient of the concentration of neutral point defects, the
drift term is added to the right hand of the equation that has the
type of Fick's second law. It means that an additional drift flux
of impurity atoms proportional to the gradient of concentration of
point defects in the neutral charge state is added to the flux
caused by an impurity concentration gradient. If these fluxes are
directed in opposition, a component of the ``uphill'' diffusion is
included into the general impurity flux. It means that segregation
of impurity atoms can be observed at great values of the gradient
of point defects in the neutral charge state instead of the
leveling effect for nonuniform impurity distribution. If a
component of the ``uphill'' diffusion exceeds the ordinary
diffusion flux described by Fick's first law, an unusual form of
the impurity profile will be observed.

As has been mentioned above, the term on the right-hand side of
the diffusion equation that describes the ``uphill'' diffusion
component has a form of a drift flux caused by a force field. As
such fields, a built-in electric field or a field of elastic
stresses can be considered. It follows from experimental data and
theoretical calculations that at a corresponding direction of the
built-in electric field the ``uphill'' impurity diffusion is
really observed \cite{Hu-68,Jones-76}. The ``uphill'' diffusion is
also observed under conditions of significant elastic stresses
that influence the diffusion of mobile species (the Gorsky effect
\cite{Gorsky-35,Geguzin-86}). However, in the case under
consideration the ``uphill'' diffusion component is due to the
so-called fictitious thermodynamic forces
\cite{DeGroot-64,Manning-71,Filippov-86} that qualitatively differ
from the real forces acting on mobile particles. As follows from
\cite{DeGroot-64,Manning-71,Filippov-86}, the fictitious
thermodynamic forces arise due to the gradient of the
concentrations of the species participating in transport
processes. In the present investigation, such species are
represented by vacancies and silicon self-interstitials being in
the neutral charge state.

In the one-dimensional case Eqs. (\ref{DifEq1}) and
(\ref{DifEq1Drift}) have the form

\begin{equation} \label{DifEq1x}
{\frac{{\partial \,C}}{{\partial \,t}}} = D_{i} \,{\frac{{\partial
^{\, 2}\left( {\tilde {C}^{\times} C} \right)}}{{\partial
\,x^{2}}}} \, ,
\end{equation}

\begin{equation} \label{DifEq1Driftx}
{\frac{{\partial \,C}}{{\partial \,t}}} = D_{i} \,{\frac{{\partial
}}{{\partial \,x}}}\left( {\tilde {C}^{\times} {\frac{{\partial
\,C}}{{\partial \,x}}}} \right) + D_{i} \,{\frac{{\partial}
}{{\partial \,x}}}\left( {{\frac{{\partial \,\tilde {C}^{\times}
}}{{\partial \,x}}}C} \right) \, .
\end{equation}

It is worth noting that the equation similar to the diffusion
equation (\ref{DifEq1x}) has been obtained in \cite{Lucas-80} for
the so-called ``kick-out'' mechanism of diffusion that is related
in general to the substitutional-interstitial diffusion mechanism
when the silicon self-interstitial displaces an immobile impurity
atom from the substitutional position to the interstitial one. A
migrating interstitial impurity atom in turn replaces the host
atom becoming substitutional again. Such analogy is not
surprising, because in later publications
\cite{Robinson-92,Velichko-08} it has been shown that there is no
difference in the mathematical description of the impurity
transports processes occurring due to ``impurity atom -- silicon
self-interstitial'' equilibrium pairs and due to the kick-out
mechanism if impurity interstitials are in local thermodynamic
equilibrium with substitutional impurity atoms and nonequilibrium
silicon self-interstitials. At the same time, there is a
substantial difference between the mathematical descriptions of
impurity transport processes due to the equilibrium
impurity-vacancy pairs and to the simple vacancy mechanism of
diffusion when the impurity atom and neighboring vacancy exchange
places. Indeed, in the papers \cite{Uskov-72,Morikawa-80} the
equation of impurity diffusion due to such a simple vacancy
mechanism has been obtained in \cite{Uskov-72,Morikawa-80}:

\begin{equation} \label{DifEqVacancy}
{\frac{{\partial \,C}}{{\partial \,t}}} = D_{i} \,{\frac{{\partial
}}{{\partial \,x}}}\left( {\tilde {C}^{V}{\frac{{\partial
\,C}}{{\partial \,x}}}} \right) - D_{i} \,{\frac{{\partial}
}{{\partial \,x}}}\left( {{\frac{{\partial \,\tilde
{C}^{V}}}{{\partial \,x}}}C} \right) \, ,
\end{equation}

\noindent where $\tilde {C}^{V}$ is the concentration of the
vacancies normalized to the thermally equilibrium vacancy
concentration $C^{V}_{eq}$.

It can be seen from Eq. (\ref{DifEqVacancy}) that the second term
in the right-hand side of this equation has a ``minus'' sign,
whereas the second term in the right hand side of Eq.
(\ref{DifEq1Driftx}) has a ``plus'' sign. It means that the flux
caused by the vacancy gradient in the simple vacancy mechanism has
an opposite direction to that for the case of impurity diffusion
due to the impurity-vacancy pairs. It follows from the analysis of
Eqs. (\ref{DifEq1Driftx}) and (\ref{DifEqVacancy}) that
investigation of impurity redistribution under conditions of
nonuniform distribution of the point defects responsible for the
impurity diffusion allows one to make a conclusion about the
character of the diffusion mechanism.

\section{Modeling of the radiation-enhanced diffusion}
\begin{figure}[!ht]
\centering {
\begin{minipage}[!ht]{9.4 cm}
{\includegraphics[scale=0.8]{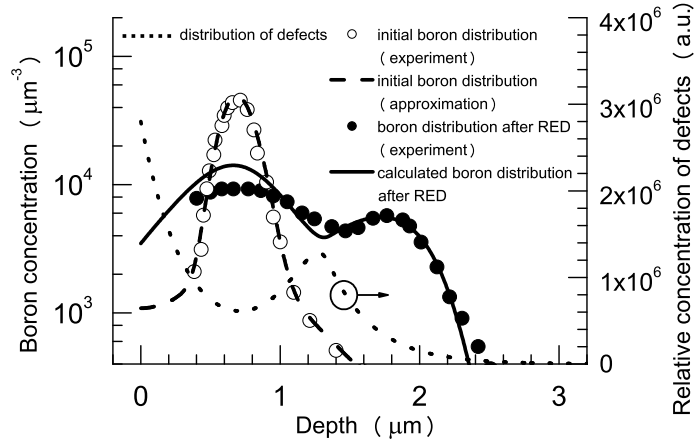}}
\end{minipage}
} \caption{Calculated boron concentration profile and distribution
of nonequilibrium point defects for radiation-enhanced diffusion
under conditions of proton implantation at a temperature of 695
$^{\circ}$C for 3 hours. Dashed and solid curves represent the
calculated distributions of boron atoms before and after the
diffusion, respectively; open and filled circles refer to the
experimental boron concentration profiles according to
\cite{Akutagawa-79}; dotted curve represents the concentration of
nonequilibrium point defects in the neutral charge state
normalized to the thermally equilibrium one. It is supposed that
the boron diffusion occurs due to the ``impurity atom -- intrinsic
point defect'' pairs} \label{fig:Akutagawa}
\end{figure}

\begin{figure}[!ht]
\centering {
\begin{minipage}[!ht]{9.4 cm}
{\includegraphics[scale=0.8]{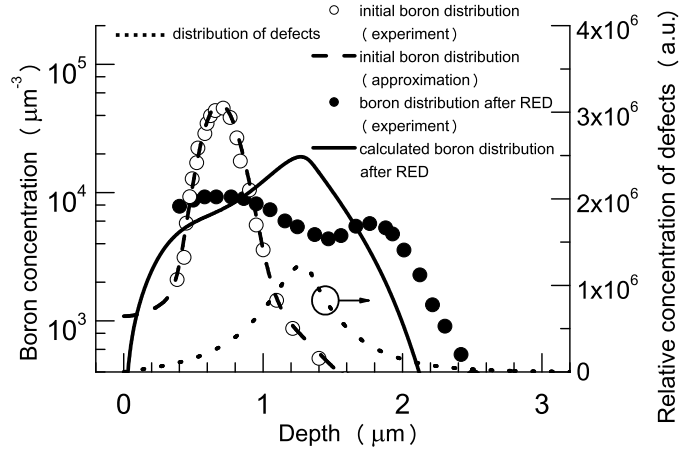}}
\end{minipage}
} \caption{Calculated boron concentration profile and distribution
of nonequilibrium point defects for radiation-enhanced diffusion
under conditions of proton implantation at a temperature of 695
$^{\circ}$C for 3 hours. Dashed and solid curves represent the
calculated distributions of boron atoms before and after the
diffusion, respectively; open and filled circles refer to the
experimental boron concentration profiles according to
\cite{Akutagawa-79}; dotted curve represents the concentration of
nonequilibrium point defects in the neutral charge state
normalized to the thermally equilibrium one. It is supposed that
boron diffusion occurs due to the simple vacancy mechanism}
\label{fig:AkutagawaVacancy}
\end{figure}

In the paper of Baruch et al. \cite{Baruch-75}, an analysis of the
experimental data obtained for the boron redistribution under
proton bombardment of homogeneously doped silicon revealed that
the impurity diffusion occurs due to the kick-out mechanism. A
similar conclusion in favor of the kick-out mechanism has been
drawn in \cite{Lucas-80,Loualiche-82} on the basis of boron
redistribution modeling in nonuniform doped silicon layers with
different maximal concentrations of impurity atoms. Modeling of
boron redistribution due to silicon bombardment by protons with an
energy of 140 keV for 3 hours at a temperature of 695 $^{\circ}$C
has been carried out in \cite{Velichko-81}. The mechanism of
impurity diffusion due to equilibrium pairs and due to the simple
vacancy mechanism was investigated. For comparison, the
experimental data obtained in \cite{Akutagawa-79} were used. In
the experiments of Akutagawa et al. \cite{Akutagawa-79} the
initial boron profile in the (111) oriented silicon substrates was
formed due to ion implantation with an energy of 300 keV and a
dose of 1.5$\times $10$^{12}$ ions/cm$^{2}$. The choice of a very
low dose was necessary for boron profiling by the differential C-V
technique to avoid the effects caused by avalanche breakdown.
Simultaneously, the effects caused by the concentration-dependent
diffusion were avoided. After the implantation and before the
proton-enhanced diffusion (140 keV protons at a beam current of
1.1 $\mu $A/cm$^{2}$), the wafers were annealed at a temperature
of 750 $^{\circ}$C for 30 min in a purified argon ambient for
electrical activation of the implanted boron. After the enhanced
diffusion, the samples are left in the target chamber at 700
$^{\circ}$C for times greater than about 30 min for postannealing
treatment, which can be expected to produce full electrical
activity of the impurity and remove the residual radiation damage.
Both the initial and the final experimental profiles of boron
concentration are presented in Figs.~\ref{fig:Akutagawa} and
~\ref{fig:AkutagawaVacancy}.

The results of modeling carried out in \cite{Velichko-81} for a
simple vacancy mechanism of boron diffusion and impurity diffusion
due to the pairs made its possible to draw a conclusion that the
radiation-enhanced diffusion under proton bombardment occurs due
to the formation, migration, and dissolution of the ``boron atom -
intrinsic point defect'' pairs that are in a local thermodynamic
equilibrium with substitutional impurity atoms and nonequilibrium
point defects. In the present, work we repeat the calculations
carried out in \cite{Velichko-81} for the case of an improved
approximation of initial impurity distribution and of the
assumption on the additional generation of point defects on the
surface of a semiconductor. Calculated boron concentration
profiles for the pair diffusion mechanism and diffusion mechanism
due to the exchange of the places between an impurity atom and a
neighboring vacancy (the simple vacancy mechanism) are presented
in Figs.~\ref{fig:Akutagawa} and ~\ref{fig:AkutagawaVacancy}. The
following values of the parameters that describe implantation of
hydrogen ions have been used for modeling: $R_{p}$ = 1.235 $\mu$m;
$\Delta R_{p}$ = 0.1124 $\mu$m; $S_{k}$ = -5.46; $R_{m}$ = 1.27
$\mu$m \cite{Burenkov-80}. Here $R_{p}$ and $\Delta R_{p}$ are the
average projective range of implanted ions and straggling of the
projective range, respectively; $S_{k}$ and $R_{m}$ are the
skewness and the position of the maximal value of the implanted
ion profile, respectively. It was supposed that the generation
rate of the nonequilibrium point defects responsible for the
diffusion of impurity atoms is proportional to the distribution of
implanted protons. The value of boron diffusivity $D_{i}$ =
3.614$\times $10$^{-11}$ $\mu $m$^{2}$/s has been calculated from
the temperature dependence presented in \cite{Pichler-04}. The
stationary distribution of nonequilibrium defects that takes into
account its additional generation on the surface of a
semiconductor was obtained from the solution of the diffusion
equation for point defects and is presented in
Fig.~\ref{fig:Akutagawa} by the dotted curve. The average
migration length of neutral point defects in intrinsic silicon
$l^{\times}_{i}$, derived from the fitting of the calculated boron
concentration profile after diffusion to the experimental one, is
equal to 0.32 $\mu$m. This value is greater than the value
$l^{\times}_{i}$ = 0.2 $\mu$m, which was used in
\cite{Velichko-81}.

It can be seen from Fig.~\ref{fig:Akutagawa} that the better
agreement of the calculated boron profile after RED with
experimental data is obtained than in \cite{Velichko-81}. Mainly,
it is due to taking into account the additional generation of
point defects on the surface of a semiconductor. Similar
calculations for the simple vacancy mechanism of impurity
diffusion are presented in Fig.~\ref{fig:AkutagawaVacancy}. As
follows from Fig.~\ref{fig:AkutagawaVacancy}, a qualitative
disagreement between the calculated boron profile and the
experimental one is observed. Thus, the present calculations
confirm the conclusion made in \cite{Velichko-81} that the
radiation-enhanced diffusion of impurity atoms during proton
bombardment of silicon substrates at elevated temperature occurs
due to the ``impurity atom -- intrinsic point defect'' equilibrium
pairs. It is worth noting that due to the exchange of its place
with the neighboring vacancy the impurity atom moves over one
interatomic distance, whereas the pair makes a great number of
jumps before the dissolution and it transfers the impurity atom to
a much larger distance. In any case, we can claim a greater
efficiency of diffusion due to the formation, migration, and
dissolution of the pairs.

Determination of the mechanism of boron diffusion during proton
bombardment allows us to investigate the form of impurity
distributions that can be obtained using the radiation-enhanced
diffusion. Let us consider, for example, the doping of silicon
from hydrogen-containing plasma with addition of some amount of a
diffusant. To provide the radiation-enhanced diffusion, it is
supposed that the substrate temperature is equal to 620
$^{\circ}$C. Such a low temperature of the substrate has been
chosen to completely exclude the thermal diffusion of impurity
atoms. Indeed, boron diffusivity calculated for this temperature
according \cite{Pichler-04} $D_{i}$ = 9.064$\times $10$^{-13}$
$\mu $m$^{2}$/s is too small for diffusion. On the other hand,
during the treatment of silicon substrates in hydrogen-containing
plasma with addition of a diffusant a great number of point
defects are generated on the surface of a semiconductor due to the
bombardment with low-energy ions. Migration of these
nonequilibrium defects into the bulk of the semiconductor provides
the radiation-enhanced diffusion of impurity atoms that enter into
the vicinity of the surface from the gas discharge plasma.

The solution of the diffusion equation for nonequilibrium neutral
point defects in the case of their generation on the surface of a
semiconductor has the form

\begin{equation} \label{ExpSolution}
\tilde {C}^{\times} \left( {x} \right) = \tilde {C}_{S}^{\times}
\,\exp \,{\left[ { - {\frac{{x}}{{l_{i}^{\times} } }}} \right]}\,
\, ,
\end{equation}

\noindent where $l_{i}^{\times}=\sqrt{d_{i} \tau_{i}}$ is the
average migration length of nonequilibrium point defects. Here
$d_{i}$ and $\tau_{i}$ are the diffusivity and average lifetime of
point defects in intrinsic silicon, respectively; $\tilde
{C}_{S}^{\times}$ is the concentration of the nonequilibrium
neutral defects on the surface of a semiconductor normalized to
the thermally equilibrium concentration of this species
${C}_{eq}^{\times}$.

\begin{figure}[!ht]
\centering {
\begin{minipage}[!ht]{9.4 cm}
{\includegraphics[scale=0.8]{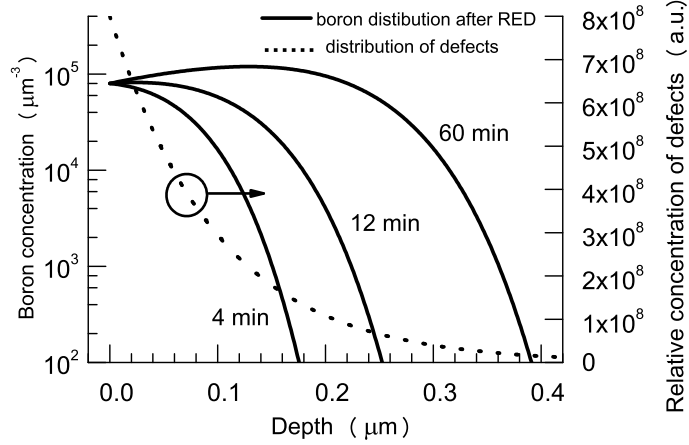}}
\end{minipage}
} \caption{Calculated boron concentration profiles for
radiation-enhanced diffusion after plasma treatment at a
temperature of 620 $^{\circ}$C for 4, 12, and 60 minutes. Dotted
curve is the concentration of nonequilibrium point defects in the
neutral charge state normalized to the thermally equilibrium one.
It is supposed that boron diffusion occurs due to the ``impurity
atom -- intrinsic point defect'' pairs} \label{fig:Plasma}
\end{figure}

The results of modeling of silicon doping with boron from the gas
discharge plasma obtained for the distribution of nonequilibrium
point defects described by expression (\ref{ExpSolution}) are
presented in Fig.~\ref{fig:Plasma}. Equation (\ref{DifEq1x})
obtained for the mechanism of impurity atom migration by the
``impurity atom -- intrinsic point defect'' pairs has been used
for the description of impurity diffusion. Taking into account the
need the formation of  shallow \textbf{\textit{ p-n}} junction,
the value of the average migration length of point defects
$l_{i}^{\times}$ = 0.1 $\mu$m has been chosen.

It can be seen from Fig.~\ref{fig:Plasma} that the impurity
profiles formed by the RED differ dramatically from the
distribution described by the \textbf{\textit{erfc}}-function
which characterizes doping from a constant source. This difference
results from the spatial dependence of the impurity diffusivity
and from the presence of the additional flux due to the
concentration gradient of point defects in the neutral charge
state. Indeed, whereas at the beginning of the treatment, the
impurity distribution is similar to the profiles formed by
diffusion with a high concentration dependence of diffusivity, the
continuation of treatment leads to the formation of a retrograde
impurity profile characterized by an increasing concentration of
impurity atoms in the bulk of a semiconductor.

Unfortunately, there is lack of experimental data on doping of
silicon from a gas discharge plasma due to the radiation-enhanced
diffusion. For example, the experimental data of \cite{Strack-63}
were obtained for the case of a long-term treatment when the
sputtering of the surface of a semiconductor plays a significant
role and, therefore, distribution of impurity atoms becomes
stationary. For the calculation of impurity distribution measured
in \cite{Strack-63}, it is reasonable to introduce a new
coordinate system $x=x^{*}-\mathrm{v_{S}}t$ bound to the moving
surface of the semiconductor and solve the diffusion equation in
this mobile coordinate system. Here $x^{*}$ is the coordinate
measured from the initial position of the surface of the
semiconductor; $\mathrm{v_{S}}$ is the projection of the surface
velocity on the immobile $x^{*}$ axis ($\mathrm{v_{S}}>0$ in the
case of sputtering of the surface of the semiconductor).

Then, the equation of impurity diffusion (\ref{DifEq1x}) in the
moving coordinate system takes the following form:

\begin{equation} \label{DifEqMovingB}
{\frac{{\partial \,C}}{{\partial \,t}}} = D_{i} \,{\frac{{\partial
\, ^{2}\left( {\,\tilde {C}^{\times} C} \right)}}{{\partial
\,x^{2}}}} + {\rm v}_{{\rm S}} {\frac{{\partial \,C}}{{\partial
\,x}}} \, \, .
\end{equation}

When the impurity atoms entering from the plasma are compensated
by those removed from the surface of the semiconductor due to
sputtering, the distribution of impurity atoms becomes stationary
in the moving system of coordinates. Then, Eq.
(\ref{DifEqMovingB}) will be transformed into a stationary
diffusion equation:

\begin{equation} \label{DifEqStationary}
D_{i} \,{\frac{{d^{\, 2}\left( {\,\tilde {C}^{\times} C}
\right)}}{{d\,x^{2}}}} + {\rm v}_{{\rm S}} {\frac{{d\,C}}{{d\,x}}}
= 0 \, \, .
\end{equation}

The ordinary differential equation (\ref{DifEqStationary}) can be
solved analytically. Let us obtain such an analytical solution for
the following Dirichlet boundary conditions:

\begin{equation} \label{BounCondDir}
C(0) = C_{S} \, , \qquad \qquad \qquad \qquad C( + \infty ) = 0 \,
\, ,
\end{equation}

\noindent where $C_{S}$ is the impurity concentration on the
surface of the semiconductor.

To obtain the analytical solution of the formulated boundary value
problem, we present Eq. (\ref{DifEqStationary}) in the following
form:

\begin{equation} \label{DifEqStatM}
{\frac{{d\,}}{{d\,x}}}{\left[ {D_{i} \,{\frac{{d\,\left( {\,\tilde
{C}^{\times} C} \right)}}{{d\,x}}} + {\rm v}_{{\rm S}} \,C}
\right]} = 0
\end{equation}

\noindent and integrate (\ref{DifEqStatM}) from $x$ up to
$+\infty$ to obtain

\begin{equation} \label{Integration}
\left. {D_{i} \,{\frac{{d\,\left( {\,\tilde {C}^{\times} C}
\right)}}{{d\,x}}}\,} \right|_{x}^{ + \infty}  + {\left. {{\rm
v}_{{\rm S}} C\,} \right|}_{x}^{ + \infty}  = 0 \, \, .
\end{equation}

As soon as both the impurity concentration and the flux of
impurity atoms are equal to zero on the right boundary
$x=+\infty$, the ordinary differential equation can be obtained
from Eq. (\ref{Integration}):

\begin{equation} \label{OrdinaryDifEq}
D_{i} \,{\frac{{d\,{\left[ {\,\tilde {C}^{\times} \left( {x}
\right)\,C\left( {x} \right)} \right]}}}{{d\,x}}}\, + {\rm
v}_{{\rm S}} \,C\left( {x} \right) = 0 \, \, .
\end{equation}

This equation can be solved using the separation of variables:

\begin{equation} \label{Separation}
{\frac{{d\,{\left[ {\,\tilde {C}^{\times} \left( {x}
\right)\,C\left( {x} \right)} \right]}}}{{\tilde {C}^{\times}
\left( {x} \right)\,C\left( {x} \right)}}} = - {\frac{{{\rm v}_{S}
\,}}{{D_{i} \,}}}{\frac{{1}}{{\tilde {C}^{\times} \left( {x}
\right)}}}d\,x \, \, .
\end{equation}

Let us integrate Eq. (\ref{Separation}) from 0 up to $x$:

\begin{equation} \label{IntegralLim}
{\left. {\ln {\left[ {\,\tilde {C}^{\times} \left( {x}
\right)\,C\left( {x} \right)} \right]}\,\,} \right|}_{0}^{x} = -
{\frac{{{\rm v}_{S} \,}}{{D_{i} \,}}}{\int\limits_{0}^{x}
{{\frac{{1}}{{\tilde {C}^{\times} \left( {x} \right)}}}d\,x}} \,
\, ,
\end{equation}

\noindent or

\begin{equation} \label{Integral}
\ln \,{\left[ {{\frac{{\,\tilde {C}^{\times} \left( {x}
\right)\,C\left( {x} \right)}}{{\,\tilde {C}_{S}^{\times} \,C_{S}}
}}} \right]} = - {\frac{{{\rm v}_{S} \,}}{{D_{i}
\,}}}{\int\limits_{0}^{x} {{\frac{{1}}{{\tilde {C}^{\times} \left(
{x} \right)}}}d\,x}} \, \, ,
\end{equation}

\noindent where $\tilde {C}_{S}^{\times}$ is the normalized
concentration of point defects in the neutral charge state on the
surface of the semiconductor.

Using exponentiation of Eq. (\ref{Integral}), one can obtain the
expression for distribution of impurity concentration:

\begin{equation} \label{GenSolution}
C\left( {x} \right) = C_{S} {\frac{{\tilde {C}_{S}^{\times}
\,}}{{\tilde {C}^{\times} \left( {x} \right)}}}\exp \,{\left[ { -
{\frac{{{\rm v}_{{\rm S}} \,}}{{D_{i} \,}}}{\int\limits_{0}^{x}
{{\frac{{1}}{{\tilde {C}^{\times }\left( {x} \right)}}}d\,x}} }
\right]} \, \, .
\end{equation}

It follows from expression (\ref{GenSolution}) that for
$\mathrm{v_{S}}=0$ the concentration of impurity atoms $C\left(
{x} \right) \rightarrow +\infty$ at $\tilde {C}^{\times}
\rightarrow 0$. At the same time for $\mathrm{v_{S}}>0$ the
impurity concentration $C\left( {x} \right)$ has a finite value at
$\tilde {C}^{\times} \rightarrow 0$ due to the faster decrease of
the exponential function.

Let us consider the widespread case of defect generation on the
surface of a semiconductor. Then, the distribution of neutral
point defects can be described by expression (\ref{ExpSolution}).
Substituting (\ref{ExpSolution}) into (\ref{GenSolution}) and
calculating the integral obtained, we find that the impurity
distribution is determined by the expression

\begin{equation} \label{PartialSolution}
C\left( {x} \right) = C_{S} \,\exp \,{\left[
{{\frac{{x}}{{l_{i}^{\times}} }}} \right]}\exp \,{\left\{ { -
{\frac{{{\rm v}_{S} \,l_{i}^{\times}} }{{D_{i} \,\tilde
{C}_{S}^{\times} } }}{\left[ {\exp \left(
{{\frac{{x}}{{l_{i}^{\times}} }}} \right) - 1} \right]}} \right\}}
\, \, .
\end{equation}

\begin{figure}[!ht]
\centering {
\begin{minipage}[!ht]{9.4 cm}
{\includegraphics[scale=0.8]{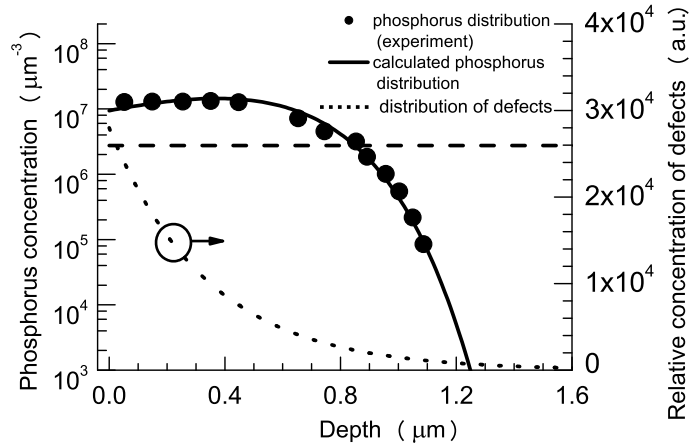}}
\end{minipage}
} \caption{Calculated phosphorus profile for radiation-enhanced
diffusion due to plasma treatment at a temperature of 820
$^{\circ}$C. Dotted curve is the concentration of nonequilibrium
point defects in the neutral charge state normalized to the
thermally equilibrium one. Filled circles are the experimental
phosphorus profile according to \cite{Strack-63}. Dashed curve is
the concentration of charge carriers $n_{i}$. It is supposed that
phosphorus diffusion occurs due to the ``impurity atom --
intrinsic point defect'' pairs} \label{fig:PlasmaSrack}
\end{figure}

It follows from expression (\ref{PartialSolution}) that the
concentration of impurity atoms increases with $x$, attains a
maximum, and then decreases to zero. The phosphorus concentration
profile calculated by means of expression (\ref{PartialSolution})
for silicon doping from the gas discharge plasma is presented in
Fig.~\ref{fig:PlasmaSrack}. The experimental data are taken from
\cite{Strack-63}. In the experiments the \textbf{\textit{p-}}type
silicon with a conductivity of 200 $\Omega \,$cm was used. The
treatment temperature was equal to 820 $^{\circ}$C and the rate of
surface sputtering was equal approximately to 5.28$\times
$10$^{-4}$ $\mu $m/s. The impurity distribution profile was found
by removing thin layers from the surface of the sample and
measuring their sheet resistance. The thermal diffusivity of
phosphorus for the above-mentioned temperature is equal to
1.919$\times $10$^{8}$ $\mu $m$^{2}$/s \cite{Haddara-00}, the
intrinsic carrier concentration, to $n_{i}$ = 2.76$\times
$10$^{6}$ $\mu $m$^{-3}$.

As can be seen from Fig.~\ref{fig:PlasmaSrack}, the impurity
concentration profile calculated from expression
(\ref{PartialSolution}) agrees well with the experimental data
obtained for phosphorus radiation-enhanced diffusion. The
agreement is observed in the entire diffusion zone including the
region near the surface of the semiconductor. To satisfy the
experimental data \cite{Strack-63}, the average migration length
of point defects $l^{\times}_{i}$ was chosen to be equal to 0.34
$\mu $m. This value is very close to the value of $l^{\times}_{i}$
= 0.32 $\mu $m, which has been derived from the experimental data
of \cite{Akutagawa-79} on the boron radiation-enhanced diffusion
during implantation of high energy hydrogen ions.

\newpage

\section{Conclusions}

To investigate the microscopic mechanisms of impurity transport in
semiconductors, modeling of boron radiation-enhanced diffusion
during implantation of high energy protons into the silicon
substrate being at an elevated temperature and modeling of
phosphorus radiation-enhanced diffusion during the treatment of
silicon substrate in a hydrogen-containing plasma with the
addition of a diffusant have been carried out. It follows from the
comparison of the calculated impurity profiles with experimental
ones that the radiation-enhanced diffusion occurs by means of
formation, migration, and dissolution of the ``impurity atom --
intrinsic point defect'' pairs which are in a local thermodynamic
equilibrium with the substitutionally dissolved impurity atoms and
nonequilibrium point defects generated due to external
irradiation. It is worth noting that a similar calculation for the
simple vacancy mechanism of diffusion due to exchange of the
places between impurity atom and neighboring vacancy qualitatively
disagrees with the experimental data.

For the pair diffusion mechanism a theoretical investigation was
carried out for the form of impurity profiles that can be created
in the vicinity of the surface due to the radiation-enhanced
diffusion during plasma treatment. It was shown that for diffusion
from a dopant source that provides a constant impurity
concentration on the surface the impurity profiles formed by the
RED are similar to those formed in the case of the high
concentration dependence of impurity diffusivity. Moreover, it is
possible to form a retrograde impurity distribution characterized
by an increasing concentration of impurity atoms in the bulk of
the semiconductor. These characteristic features of the dopant
profile are due to the nonuniform distribution of neutral point
defects responsible for the impurity diffusion.

For simulation of the experimental phosphorus profile
\cite{Strack-63} formed during plasma treatment, the analytical
solution of the equation that describes the radiation-enhanced
diffusion under conditions of sputtering of silicon surface was
obtained. From fitting to the experimental data the average
migration length of point defects $l^{\times}_{i}$ = 0.34 $\mu $m
was obtained. This value is very close to $l^{\times}_{i}$ = 0.32
$\mu $m which was derived from the experimental data of
\cite{Akutagawa-79} on the boron radiation-enhanced diffusion
during ``hot'' implantation of high energy protons.

The results of the calculations performed give a clear evidence in
favor of further investigation of various doping processes based
on the radiation-enhanced diffusion, especially the processes of
plasma doping, to develop a cheap method for formation of special
impurity distributions in the near-surface region.

\end{document}